\documentclass[12pt]{article}
\usepackage{graphicx}
\title{Tests of Gaussianity}

\date{}

\author{A.M. Aliaga\footnote{Instituto de F\'{\i}sica de Cantabria, CSIC--Univ.
de Cantabria, Avda. Los Castros s/n, 39005 Santander, Spain} \footnote{Dpto. de
F\'{\i}sica Moderna. Univ. de Cantabria, Avda. Los Castros s/n, 39005 
Santander,Spain} , 
E. Mart\'{\i}nez--Gonz\'alez$^*$, 
L. Cay\'on$^*$\footnote{Physics Department, Purdue University, 525 
Northwestern Avenue, West Lafayette, IN 47907-2036, USA} ,\\
F. Arg\"ueso\footnote{Dpto. de Matem\'aticas, Univ. de Oviedo, C/ Calvo Sotelo 
s/n, 33007 Oviedo, Spain} , 
J.L. Sanz$^*$, 
R.B. Barreiro$^*$,
J.E. Gallegos$^*$
}

\begin{document}
\maketitle

\begin{abstract}
We review two powerful methods to test the Gaussianity of the cosmic
microwave background (CMB): one based on the distribution of spherical
wavelet coefficients and the other on smooth tests of
goodness-of-fit. The spherical wavelet families proposed to analyse the
CMB are the Haar and the Mexican Hat ones. The latter is preferred for
detecting non-Gaussian homogeneous and isotropic primordial models
containing some amount of skewness or kurtosis. Smooth tests of
goodness-of-fit have recently been introduced in the field showing
some interesting properties. We will discuss the smooth tests of 
goodness-of-fit developed by Rayner and Best for the univariate as 
well as for the multivariate analysis.

\end{abstract}

\section{Introduction}

Establishing the statistical character of the CMB fluctuations is one
of the most fundamental problems in cosmology. The simplest
inflationary theories predict Gaussian fluctuations whereas
non-standard inflation and topological defects predict different
non-Gaussian ones. Recent sensitive CMB observations (Boomerang,
MAXIMA, DASI, WMAP) have shown no evidence of departures from Gaussianity up
to date. This fact has put strong constraints on the amount of cosmic
strings in the universe and on the non-linear coupling parameter in
the case of a quadratic potential (Spergel et al. 2003).

There is not a unique way to search for non-Gaussianity in the CMB. Different 
features will be best probed by methods which are well adapted to point them.
The methods can work in real space and also in Fourier, wavelet or other 
spaces in which the non-Gaussian features can be more enhanced. In
this work we will focus on two powerful methods to search for
non-Gaussianity: one based on spherical wavelets and the other on 
smooth goodness-of-fit tests. Only two spherical wavelet families 
have been considered in CMB analyses: Haar and Mexican Hat. Some of 
their properties as well as their potentiality to test the normality 
will be discussed. The smooth tests of goodness-of-fit have been shown 
to be very powerful in testing the Gaussianity of data in many fields
outside the CMB. We here introduce the tests in the CMB context and
explain how they can be applied in a satisfactory way. We will
first consider the univariate case, already introduced in Cay\'on et
al. (2003b), and later the multivariate one. The problematics
concerning their application to the CMB sky will be discussed.  
We will also review some recent results obtained by applying the
previous methods to different experiments like COBE-DMR and MAXIMA.

\section{Spherical wavelets}

Although several works have already been performed using planar
wavelets to analyse the CMB temperature fluctuations in small patches
of the sky, it is clear that a proper analysis of any area of the sky
will involve spherical wavelets. The latter have been developed recently. 
Schr{\"o}der and Sweldens (1995) introduced the spherical Haar wavelets 
(SHW) as a generalization of the planar ones to the pixelised sphere. They 
are orthogonal and adapted to a given pixelization of the sky which 
must be hierarchical. Two applications of the spherical Haar wavelets 
have already been performed to analyse CMB maps. Tenorio et al. (1999) 
applied them to a QuadCube pixelization of the CMB sky for which a 
correction of the pixel area has to be made. Barreiro et al. (2000)
tested the Gaussianity of the COBE-DMR data on the HEALPix equal-area
pixelation for which no correction is required. They found COBE-DMR
data consistent with Gaussianity.

An extension of isotropic wavelets to the sphere has been proposed by
Antoine and Vandergheynst (1998) following a group theory approach.
It is based on the stereographic projection which translates from the
plane to the sphere the following four basic properties:
compensation, translation, dilation and the Euclidean limit for small
angles. The spherical Mexican Hat wavelets (SMHW) are a particular case of
isotropic wavelets which can thus be constructed by a stereographic
projection of the planar Mexican Hat ones (Mart\'\i nez-Gonz\'alez et
al. 2002). Cay\'on et al. (2001) tested the Gaussianity of the COBE-DMR 
data on the HEALPix pixelation using the skewness and kurtosis of the 
SMHW cofficients  at different angular scales and also the correlations 
at different scales. No deviation from Gaussianity was detected.

A comparative analysis of the performance of the two spherical wavelet
families, SHW and SMHW, already proposed in the literature to analyse the CMB
statistical distribution has been performed by Mart\'\i nez-Gonz\'alez et
al. (2002). The comparison was based on the power to discriminate between 
standard inflationary models (Gaussian) and non-Gaussian models, the 
latter consisting in small deviations of the Gaussian inflationary models 
introducing an artificially specified skewness or kurtosis through the 
Edgeworth expansion. The skewness or kurtosis present at different wavelet 
scales was included by a linear optimal combination using the Fisher 
discriminant. The result of this analysis is that the SMHW is much more 
sensitive to these low-order cumulants than the SHW and it is therefore 
preferred for the detection of this type of non-Gaussianity.

More recently, the nonlinear coupling parameter $f_{nl}$ in the case
of a quadratic term for the gravitational potential has been
constrained with the COBE-DMR data (Cay\'on et al. 2003a). 
The result, $f_{nl}<1100$ ($68 \%$ confidence level) represents the strongest
constraint obtained with those data.  

\section{Smooth tests of goodness-of-fit}

In this section, the \emph{score} statistic is presented. This statistic is the
starting point to construct the Rayner and Best test (Rayner and Best 1989, 
1990), which is used in the present work to test the univariate and 
multivariate Gaussian distributions.

Given a statistical variable $x$ and $n$ independent realizations $\{ x_i\}$ ($i=1, \ldots, n$), we want to test if the probability density function of $x$ is equal to $f(x)$. This statistical variable can be a real $p$-dimensional vector and, thus, the method can test, for example, if $f$ is a multivariate normal distribution (this is one of the cases we are interested in). Smooth tests are constructed to discriminate between the predetermined function $f(x)$ and a second one that deviates smoothly from the former. We consider an alternative probability density function $f(x,\theta)$ (where $\theta$ is a parameter vector) that deviates smoothly from $f(x)$ and $f(x,\theta_0) = f(x)$. In other words, we want to test the \emph{null hypothesis}: $\theta = \theta_0$.

The probability density of $n$ independent realizations $\{ x_i\}$ is given by $\prod_{i=1}^{n} f(x_i,\theta)$. Given these measurements we calculate the estimated value of $\theta$ by means of the \emph{Maximum Likelihood Method}. We denote this value by $\hat{\theta}$.

We define $W$ such that
\begin{equation}
\exp \bigg( \frac{W}{2} \bigg) = \frac{\prod_{i=1}^{n} f(x_i,\hat{\theta})}{\prod_{i=1}^{n} f(x_i,\theta_0)} 
\end{equation}

Let $U(\theta)$ be a vector whose components are $U_i(\theta) \equiv \partial \ell(\{ x_j\}, \theta) / \partial \theta_i$, where the log--likelihood is defined as $\ell (\{ x_j\}, \theta) \equiv \log \prod_{i=1}^{n} f(x_i,\theta)= \sum_{i=1}^{n} \log f(x_i,\theta)$.

Assuming $\hat{\theta}$ close to $\theta_0$, we expand $W$ and $U(\theta_0)$ in Taylor series around $\hat{\theta}$, and we have

\begin{equation}
W \approx - U^T(\theta_0) \bigg\{ \frac{\partial U(\theta_0)}{\partial \theta} \bigg\}^{-1} U(\theta_0)
\end{equation}

This quantity is a direct measurement of the diference between $\hat{\theta}$ and $\theta_0$ and therefore a test of the null hypothesis. We construct a quantity closely related to the last one. To construct this quantity we substitute $-\partial U(\theta_0) / \partial \theta$ for its mean value

\begin{equation}
S = U^T(\theta_0) I^{-1}(\theta_0) U(\theta_0)
\end{equation}

\noindent where $I$ is a matrix of components $I_{ij}(\theta) \equiv \langle U_i(\theta)U_j(\theta) \rangle$ and it can be shown that it is equal to $- \langle \partial U_i(\theta) / \partial \theta_j \rangle$. The approximation of $- \partial U_i(\theta) / \partial \theta_j$ by its mean value is valid when $n$ is large\footnote{
In fact
$I_{ik}(\theta) = - \langle \partial U_i(\theta)/ \partial \theta_k \rangle = - \sum _{j=1}^n \langle \partial^2 \log f(x_j,\theta) / \partial \theta_k \partial \theta_i \rangle= - n \langle \partial^2 \log f(x_j,\theta) / \partial \theta_k \partial \theta_i \rangle$, where the last equality holds because $x_j$ is the same statistical variable for all $j$. When $n$ is large, the last mean value can be approximated by  $(1/n)\sum_{j=1}^n \partial^2 \log f(x_j,\theta) / \partial \theta_k \partial \theta_i$, and then, $I_{ik}(\theta) \approx - \partial U_i(\theta)/ \partial \theta_k $ when $n \rightarrow \infty$.
}.
The $S$ quantity is the so--called \emph{score} statistic. A wider description of the previous development can be found in Cox and Hinkley (1974).

\subsection{Rayner-Best test: univariate Gaussian}

Let us suppose that we have a statistical variable which takes values in the real domain. In the work of Rayner and Best (1989,1990) it is defined an alternative probability density function of orden $k$ given by

\begin{equation} \label{eq:001}
g_k(x)= C(\theta_1,\ldots,\theta_k) \exp \bigg\{ \sum_{i=1}^k \theta_i h_i(x) \bigg\} f(x)
\end{equation}

\noindent where $C$ is a normalization constant and the $h_i$ functions are orthonormal on $f$ with $h_0(x)=1$. Then, the score statistic associated to the $k$ alternative is given by

\begin{equation}\label{eq:002}
S_k = \sum_{i=1}^k U_i^2 \qquad \textrm{with} \qquad U_i=\frac{1}{\sqrt n} \sum_{j=1}^n h_i(x_j)
\end{equation}

When we want to test if $f$ is a Gaussian function of zero mean and unit variance, the $h_i$ functions become the normalized Hermite--Chebishev polynomials, that is, they are equal to $P_n(x)/s_n$ with $s_n=\sqrt{n!}$ and $P_0(x)=1$, $P_1(x)=x$ and for $n \ge 1$: $P_{n+1}(x)=xP_n(x)-nP_{n-1}(x)$. 

Every alternative function, that is, every $k$ value, gives a statistic $S_k$. The statistics $S_k$ are given by: $S_1  = n (\hat{\mu}_1)^2$, $S_2  = S_1 + n (\hat{\mu}_2 - 1 )^2 /2 $, $S_3  = S_2 + n (\hat{\mu}_3 - 3 \hat{\mu}_1 )^2 /6$, $S_4  =  S_3 + n ( [\hat{\mu}_4 - 3 ] - 6 [\hat{\mu}_2 -1] )^2  /24 $  and $S_5  =  S_4 + n ( \hat{\mu}_5 - 10 \hat{\mu}_3 + 15 \hat{\mu}_1)^2 /120 $, where $\hat{\mu}_{\alpha}=(\sum_{j=1}^n x_j^{\alpha})/n$. Thus, the statistic $S_k$ is related to cumulants of order $\le k$, and then this test is directional, that is, it indicates how the actual distribution deviates from Gaussianity. For example, if $S_1$ and $S_2$ are small and $S_3$ is large, then the data have a large $\hat{\mu}_3$ value and also have a large skewness value because of the relation between $\hat{\mu}_3$ and $S_3$.

When $n \rightarrow \infty$, the $S_k$ statistic is distributed as a $\chi_k^2$. This holds because $U_i$ is Gaussian distributed when $n \rightarrow \infty$ (sum of a large number of independent variables).

In the case of the CMB, the data are correlated and therefore not independent. Let $x_i$ be the value of the pixel $i$. If we perform the Cholesky decomposition of the correlation matrix of the data: $C=LL^T$, and change to the variable $y_j=\sum_i L_{ji}^{-1}x_i$, then these new data are uncorrelated with zero mean and unity deviation and then, if Gaussianity holds, they are independent with a $N(0,1)$ distribution. Then we work with these data and apply to them the test here described. 

In Cay\'on et al. (2003b), the power of these statistics is studied and the method is used to test the Gaussianity of the data of the MAXIMA experiment (Balbi et al. 2000, Hanany et al. 2000). The analysis finds these data compatible with Gaussianity under these goodness-of-fit tests.

These statistics are also used in Aliaga et al. (2003) to constrain the skewness and kurtosis of the mentioned data. In this work the skewness $S$ and the kurtosis $K$ are constrained to $|S|\le 0.035$ and $|K|\le 0.036$, at 99\% confidence level. 

\subsection{Rayner-Best test: multivariate Gaussian} \label{rbmul}

In this case, the statistical variable $x$ takes values in a $p$-dimensional vectorial space and we want to test if $f(x)$ is a multinormal distribution of mean equal to $\mu$ and correlation matrix equal to $\Sigma$. For the sake of clarity we are going to change the notation of the vectorial quantities $x$ and $\mu$ for $\vec x$ and $\vec \mu$. As it happened in the previous subsection, we need a set of orthonormal functions on $f$. Rayner and Best (1989) propose the following construction: the Cholesky decomposition of the correlation matrix is performed: $\Sigma=LL^T$ and the matrix $A \equiv L^{-1}$ is obtained. Thus, one constructs the new variables: $\vec y=A(\vec x-\vec \mu)$. It is straightforward to see that the functions $L_{r_1 \cdots r_p}(\vec y)= H_{r_1}(y_1) \cdots H_{r_p}(y_p)$ are orthonormal on $f(\vec x)$, where the $H_s$ function is the normalized Hermite-Chebishev polynomial of degree $s$ and $y_r$ ($r=1,\ldots,p$) is the $r$ component of the $\vec y$ vector. The degree of the $L_{r_1 \cdots r_p}$ function is equal to $ r = r_1+ \cdots +r_p$.

In the case of the Rayner-Best test applied to the univariate Gaussian distribution, the $h_i$ functions are ordered by its degree $i$. In that case the ordering is easy to establish, but in the case of the multivariate Gaussian, different $L_{r_1 \cdots r_p}$ functions can have the same degree. Thus, suppose some ordering has been imposed, such that the degree $r$ functions are considered before the degree r+1 ones. Call this ordered system $\{ L_s^{(r)}(y)\}$, where $s$ is the position in the ordering and $r$ indicates the degree of the function.

It holds that the number of functions of degree $r$ is the combinatorial number $^{p+r-1}C_r$.

To calculate the $S_k$ statistic, we start from an alternative distribution like the one shown in the expresion (\ref{eq:001}), but the $h_s$ functions are substituted by the $L_s^{(r)}$ ones. The order of the alternative function, that is, the index $k$ of $S_k$ represents the number of functions $L_s^{(r)}$ used. If we use functions up to degree $r$, then $k= \sum_{s=1}^r {}^{p+s-1}C_s = {}^{p+r}C_r-1$. In an analogous way to the univariate case, the $S_k$ is given by

\begin{equation} \label{eq:003}
S_k = \sum_{i=1}^k V_i^2 \qquad \textrm{with} \qquad V_i=\frac{1}{\sqrt n} \sum_{j=1}^n L_i^{(s)}(A(\vec x_j-\vec \mu))
\end{equation}

It is interesting to construct quantities like the $U_i^2$ ones of the expresion (\ref{eq:002}) which have a defined degree and their addition gives the $S_k$ statistic. To do this, in the sum of the $V_i^2$ terms in (\ref{eq:003}) one separates and groups the terms which have the same degree, and then, if we go up to degree $r$:

\begin{equation} \label{eq:004}
S_k = U_1^2 + \cdots + U_r^2 \qquad \textrm{with} \qquad U_s^2= \frac{1}{n}\sum_i  \bigg[ \sum_{j=1}^n L_i^{(s)}(A(\vec x_j-\vec \mu)) \bigg]^2
\end{equation}

In the previous expresion, for every $U_s^2$, the sum on the $i$ index is made over all the functions of degree equal to $s$ and so $U_s^2$ is a quantity of degree $s$. The $\vec x_j$ vector is the sample $j$ of our experiment. When $n \rightarrow \infty$, $V_i$ is Gaussian distributed (with zero mean and unit variance) because of the Central Limit Theorem applied to the independent $\{\vec x_j\}$ samples. Thus $V_i^2 \sim \chi^2_1$ and $U_s^2 \sim \chi^2_{\nu}$ with $\nu = {}^{p+s-1}C_s$. If we calculate the $S_k$ statistics, or the $U_s^2$ quantities, as it is shown in equation (\ref{eq:004}), it is not important the order we have imposed to the $L_i^{(s)}$ functions, the only important aspect is how many functions of degree $s$ are there. If we know these functions we only have to add them to construct $U_s^2$.

In the appendix, the $U_s^2$ quantities are given, for $s=1,2,3,4$, and their distributions are shown in Figure \ref{disu2}. These distributions are calculated with 5000 realizations where one realization consists of $n=5000$ 3-dimensional vectors ($p=3$). Then, as $n$ is very large, we see that the shown distributions are very close to the asymptotic distributions, that is, $U_1^2 \sim \chi^2_3$, $U_2^2 \sim \chi^2_6$, $U_3^2 \sim \chi^2_{10}$ and $U_4^2 \sim \chi^2_{15}$. In the Table \ref{tab1}, the values of the mean and standard deviation ($\sigma$) of the $U_s^2$ distributions are shown. These values are compared with the values of the corresponding asymptotic distributions $\chi^2_{\nu}$. The distributions of the $S_3$ and $S_4$ statistics are shown in the Figure \ref{dis_s}. For $S_3$ the obtained mean value is 18.961 and the standard deviation 6.259. For $S_4$ the same quantities are 33.932 and 8.534. In the asymptotic case $S_3 \sim \chi^2_{19}$ (with mean value 19 and standard deviation 6.164) and $S_4 \sim \chi^2_{34}$ (with mean value 34 and standard deviation 8.246).

As an example, we calculate the distributions of the statistics when the $y_{jr}$ are uniformly distributed with mean zero and standard deviation equal to one. The first moment of this distribution which is different from the moments of the Gaussian distribution is the kurtosis, that is, the moment of degree four. So, the distribution for the statistic $U_4^2$ must be different (see equation (\ref{eq:006})) when we compare the uniform case with the Gaussian one. The mean value and the standard deviation of $U_s^2$ distributions are shown in Table \ref{tab2}, and in fact, we see that we can discriminate between a Gaussian distribution and a uniform distribution (in this case with $p=3$), because the distributions of $U_4^2$ are very different. Also the distributions of $U_3^2$ are different enough for discrimination in this case.

Finally, the question is how to apply this test when we have only one map of CMB. We need a large number of samples $n$ to estimate the sums from $j=1$ to $n$ that we have in expressions (\ref{eq:005}) to (\ref{eq:006}), that is, given the initial map we need to extract independent samples. One way to obtain them is: first, to define the dimension $p$ of the multinormality we want to test and then divide the CMB map in patches of $p$ pixels distributed in the same way within each path; second, to try to decorrelate them. Assuming that all the data are multinormal then the patches can be considered as independent samples after decorrelation. In this way we can apply the method described above, work is in progress to apply it to WMAP.

\vspace{0.8cm}
\emph{Acknowledgements.} We acknowledge financial support provided by the Spanish Ministerio de Ciencia y Tecnolog\'{\i}a, project ESP2002-04141-C03-01. 


\section*{Appendix}
In this appendix the $U_s^2$ quantities for the multinormal distribution (section \ref{rbmul}) are explicitly given for $s=1,2,3,4$.We consider an experiment consisting in $n$ $p$-dimensional vectors, where the $r$ component of the vector $j$ is denoted by $y_{jr}$ ($r=1, \ldots, p$). From equation (\ref{eq:004}) and taking the explicit forms of the normalized Hermite--Chebishev functions we can obtain the following expresions:

\begin{eqnarray} \label{eqn:1}
U_1^2 &= &\frac{1}{n} \sum_{r=1}^p \bigg \{ \sum_{j=1}^n y_{jr} \bigg\}^2 \label{eq:005}
\\
U_2^2 &= &\frac{1}{n} \bigg[ \frac{1}{2}\sum_{r=1}^p \bigg \{ \sum_{j=1}^n (y_{jr}^2-1) \bigg\}^2
+\sum_{r=1}^{p-1}\sum_{s=r+1}^p \bigg\{ \sum_{j=1}^n y_{jr} y_{js} \bigg\}^2\bigg]
\\
U_3^2 &= &\frac{1}{n} \bigg[ \frac{1}{6}\sum_{r=1}^p \bigg \{ \sum_{j=1}^n (y_{jr}^3-3y_{jr}) \bigg\}^2
+\frac{1}{2}\sum_{r=1}^{p}\sum_{\mathop{s=1} \limits_{(s\neq r)}}^p \bigg\{ \sum_{j=1}^n (y_{jr}^2-1) y_{js} \bigg\}^2 + \nonumber \\
&& \sum_{r=1}^{p-2}\sum_{s=r+1}^{p-1}\sum_{t=s+1}^{p}\bigg\{ \sum_{j=1}^n y_{jr} y_{js} y_{jt}\bigg\}^2
\bigg]
\\
U_4^2 &= &\frac{1}{n} \bigg[\frac{1}{24}\sum_{r=1}^p \bigg \{ \sum_{j=1}^n (y_{jr}^4-6y_{jr}^2 + 3) \bigg\}^2
+\frac{1}{6}\sum_{r=1}^{p}\sum_{\mathop{s=1} \limits_{(s\neq r)}}^p \bigg\{ \sum_{j=1}^n (y_{jr}^3-3y_{jr}) y_{js} \bigg\}^2 + \nonumber
\\
&&\frac{1}{4}\sum_{r=1}^{p-1}\sum_{s=r+1}^p \bigg\{ \sum_{j=1}^n (y_{jr}^2-1)(y_{js}^2-1) \bigg\}^2+
\frac{1}{2}\sum_{r=1}^{p}\sum_{\mathop{s=1} \limits_{(s\neq r)}}^{p-1}\sum_{\mathop{t=s+1} \limits_{(t\neq r)}}^{p}  \bigg\{ \sum_{j=1}^n (y_{jr}^2-1) y_{js}y_{jt} \bigg\}^2 + \nonumber 
\\
&&\sum_{r=1}^{p-3}\sum_{s=r+1}^{p-2}\sum_{t=s+1}^{p-1}\sum_{q=t+1}^{p} \bigg\{ \sum_{j=1}^n y_{jr} y_{js} y_{jt} y_{jq}\bigg\}^2
\bigg] \label{eq:006}
\end{eqnarray}

\begin{figure}
\begin{center}
\includegraphics[width=8cm,height=6cm]{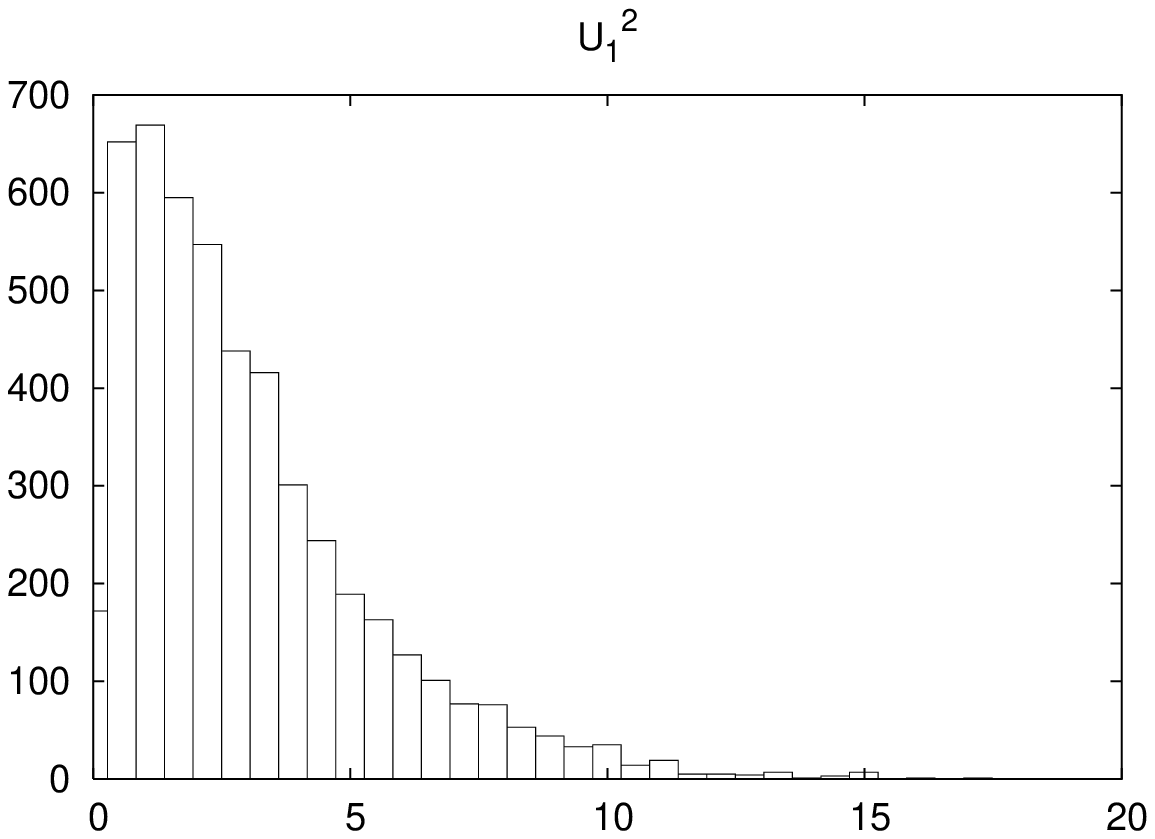}
\includegraphics[width=8cm,height=6cm]{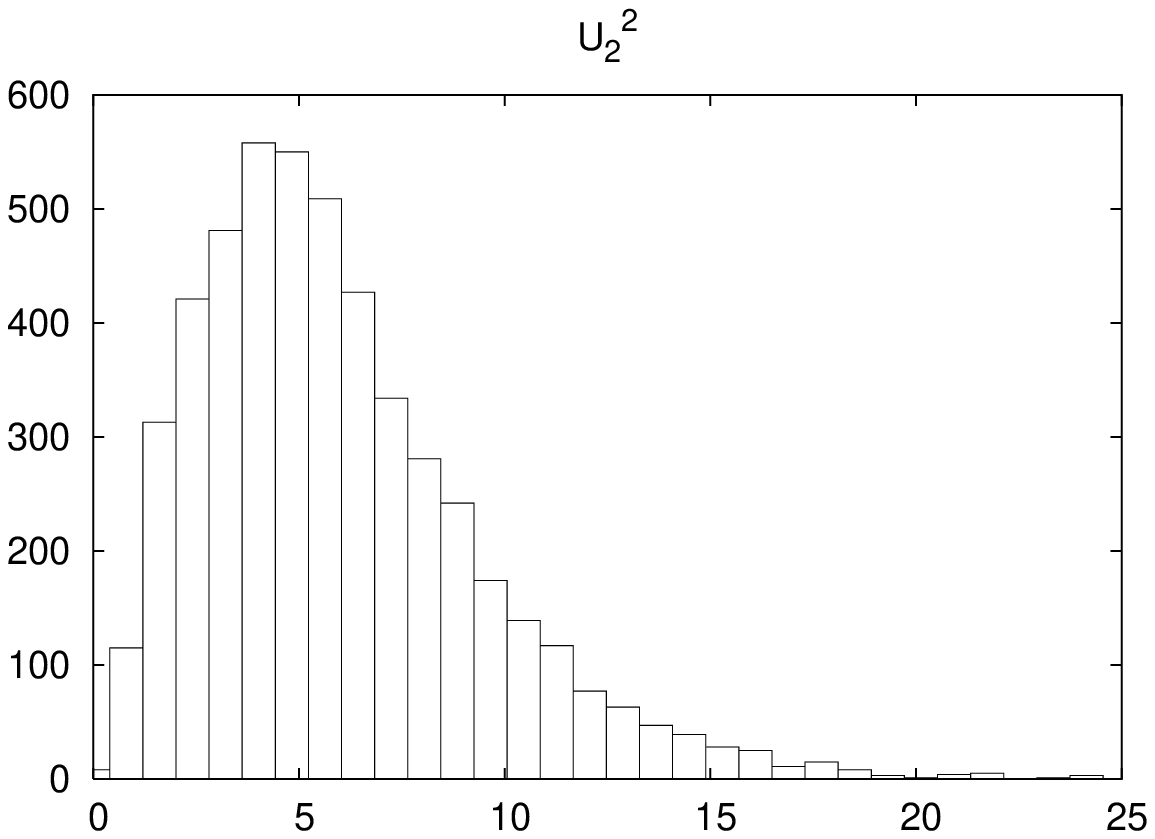}

\includegraphics[width=8cm,height=6cm]{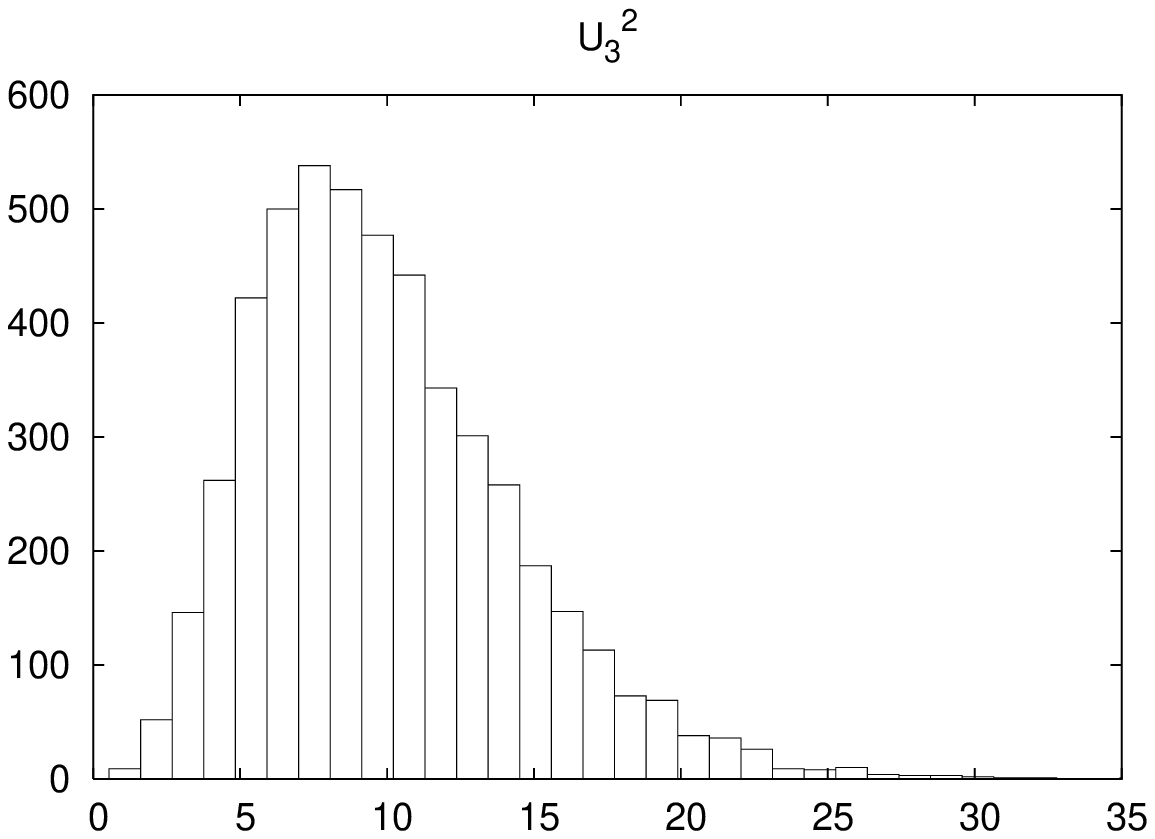}
\includegraphics[width=8cm,height=6cm]{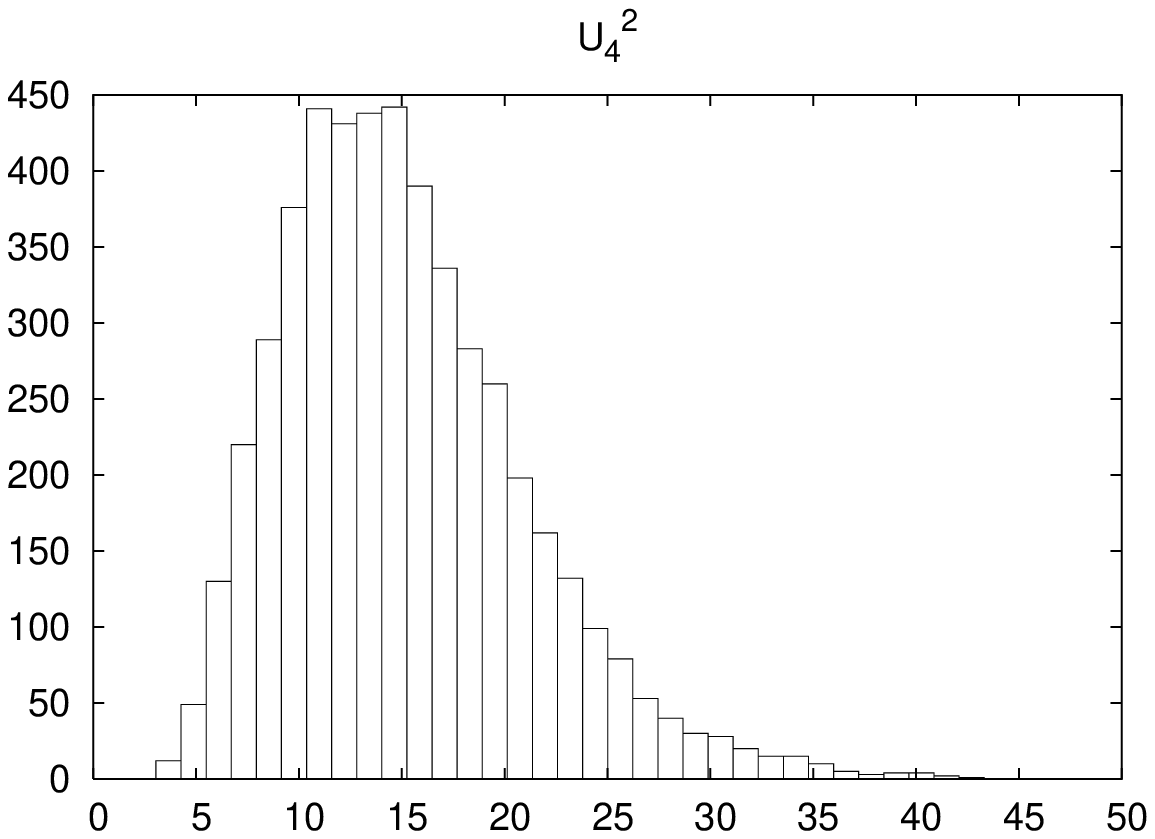}
\caption{The $U_s^2$ distributions obtained from 5000 realizations of experiments of $n=5000$ vectors of dimension $p=3$. From left to right, top to bottom $s=1,2,3,4$.}
\label{disu2}
\end{center}
\end{figure}

\begin{figure}
\begin{center}
\includegraphics[width=8cm,height=6cm]{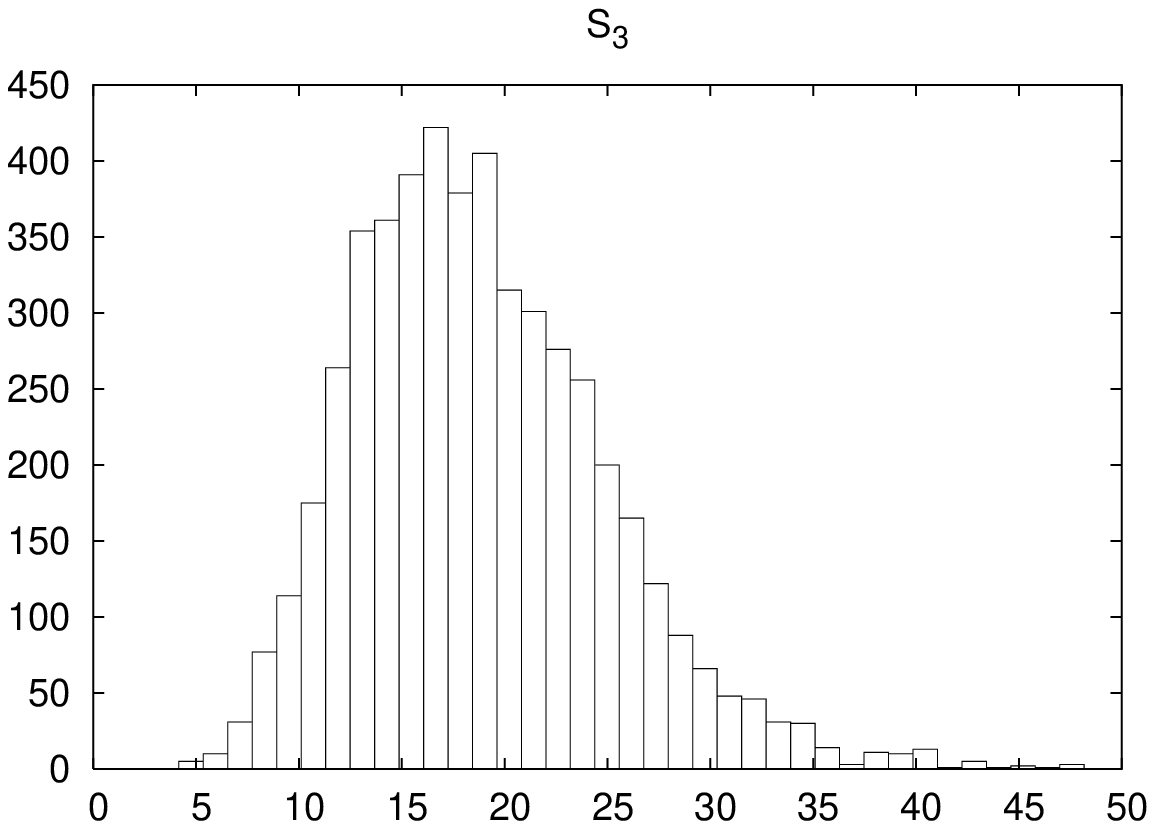}
\includegraphics[width=8cm,height=6cm]{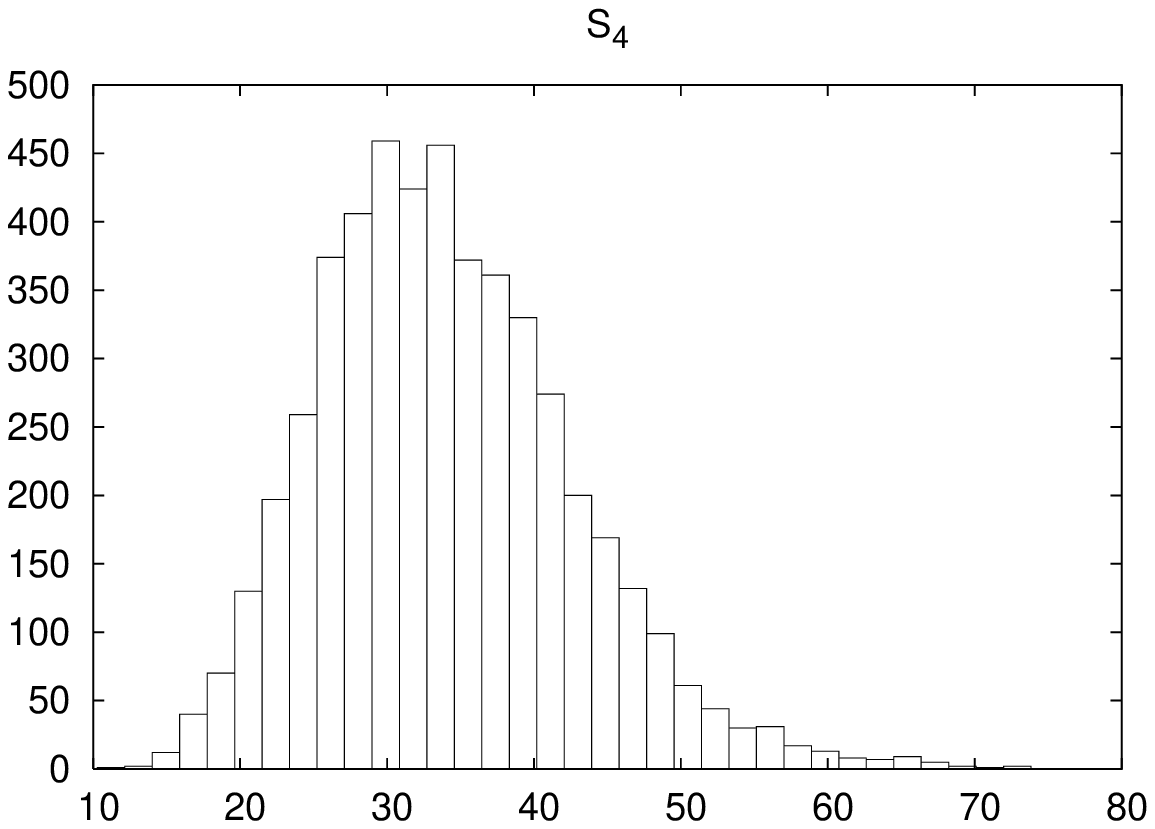}

\caption{The $S_3$ and $S_4$ distributions obtained from 5000 realizations of experiments of $n=5000$ vectors of dimension $p=3$.}
\label{dis_s}
\end{center}
\end{figure}

In order to establish the validity of the multinormal hypothesis we need to know the distribution for each of the statistics $U_s^2$ and also their combination to get $S_k$. With these distributions we can know if a given data set si compatible with multinormality. The $U_s^2$ distributions for the case $n=5000$ and $p=3$ are shown in Figure \ref{disu2} and Table \ref{tab1} compares the values of the mean and the standard deviation ($\sigma$) obtained for 5000 realizations with the asymptotic values. The mean values of all $U_s^2$ are equal to $\nu$ independently of the $n$ value ($\nu = {}^{p+s-1}C_s$). The distributions of $S_3$ and $S_4$ are shown in Figure \ref{dis_s}. 

\begin{table}
\begin{center}
\begin{tabular}{|c| c c| c c| c c| c c|}
\hline
      &$U_1^2$&$\chi^2_3$&$U_2^2$&$\chi^2_6$&$U_3^2$&$\chi^2_{10}$&$U_4^2$&$\chi^2_{15}$ \cr
\hline
   mean  & 2.966 &   3.000  &  5.968&   6.000  & 10.031&   10.000    & 15.012&   15.000   \cr

$\sigma$ & 2.480&    2.449  &  3.462&   3.464  &  4.551&    4.472    &  5.763&    5.477   \cr
\hline
\end{tabular}
\caption{Values of the mean and standard deviation ($\sigma$) of the distributions of $U_s^2$ for 5000 realizations of an experiment consisiting in $n=5000$ samples of a data vector of dimension $p=3$ multinormally distributed whose components are independent. These values are compared with the values of the corresponding asymptotic distributions $\chi^2_{\nu}$. The relation between $p$, $s$ and $\nu$ is $\nu = {}^{p+s-1}C_s$.}\label{tab1}
\end{center}
\end{table}

\begin{table}
\begin{center}
\begin{tabular}{|c| c c| c c| c c| c c|}
\hline
      &$U_1^2$(g)&$U_1^2$(u)&$U_2^2$(g)&$U_2^2$(u)&$U_3^2$(g)&$U_3^2$(u)&$U_4^2$(g)&$U_4^2$(u) \cr
\hline
   mean  & 2.966 & 2.975   &  5.968& 4.192    & 10.031& 4.353    & 15.012&  904.9  \cr

$\sigma$ & 2.480 & 2.428   &  3.462& 2.601    &  4.551& 2.140    &  5.763&  38.67   \cr
\hline
\end{tabular}
\caption{Values of the mean and standard deviation ($\sigma$) of the distributions of $U_s^2$ for 5000 realizations of an experiment consisting in $n=5000$ samples of a data vector of dimension $p=3$ multinormally distributed (g) and uniformly distributed (u) (each component is uniformly distributed). The components of the data vectors have zero mean, unit variance and are uncorrelated.}\label{tab2}
\end{center}
\end{table}

\end{document}